\documentstyle[11pt,Cargesepasp,twoside,epsf]{article}
\def\edcomment#1{\iffalse\marginpar{\raggedright\sl#1\/}\else\relax\fi}
\reversemarginpar

\begin{document}
\title{The Origin of The Stellar Initial Mass Function}
 \author{Bruce G. Elmegreen}
\affil{IBM Research Division, T.J. Watson Research Center, P.O. Box 218,
Yorktown Hts, NY 10598 USA, bge@watson.ibm.com}

\begin{abstract}
Observations and theories of the stellar initial mass function are
reviewed. The universality and large total mass range of the power law
portion of the IMF suggest a connection between stellar mass and cloud
structure.  The fractal patterns and formation times of young clusters
suggest a further connection to turbulence.  There is also a similarity
between the characteristic mass at the lower end of the power law and
the minimum self-gravitating mass in a typical star-forming cloud.
All of this suggests that the power law part of the IMF comes from
star formation in self-gravitating cloud pieces that are formed by
compressible turbulence.  Timing constraints involving cloud destruction
and competition for gas might limit the upper stellar mass to several
hundred suns. This upper limit has to be less than the mass of a clump
that has a dynamical time equal to several times the dynamical time of
the characteristic mass.  The smallest stars and brown dwarfs may not
come directly from cloud clumps, which are not self-gravitating at this
mass in the cloudy phase, but from collapsed fragments or other pieces
connected with the formation of more massive stars. \end{abstract}


\section{Introduction}

The stellar initial mass function (IMF) is an important ingredient for
studies of stellar evolution in clusters and galaxies, and it offers a
clue to the physical processes involved with star formation. But since the
time when the first IMF was extracted from field stars corrected for age
(Salpeter 1955), there has been no general understanding of its origin,
and few direct observations of stellar mass functions in their initial
form, i.e., in very young clusters.

Fortunately, the observational situation is improving with infrared
surveys down to sub-stellar masses in nearby young clusters (Comer\'on,
et al. 1993; Strom, Kepner, \& Strom 1995; Lada \& Lada 1995; Luhman,
et al. 1998; Luhman \& Reike 1998, 1999; Najita, Tiede \& Carr 2000) and
with ground- and space-based observations of distant massive clusters
(e.g., Massey \& Hunter 1998; Grillmair et al. 1998; Selman et al.
1999; Sirianni et al. 2000; see reviews in Massey 1998 and Brown 1998).

The theory of the IMF is making progress too. There are numerical
simulations that reproduce the IMF by following the motion and accretion
of protostars in a model cluster (Bonnell et al. 1997) or the evolution
of protostar interactions (Price \& Podsiadlowski 1995), and there are
simulations based on sampling from fractal clouds (Elmegreen 1997, 1999a,
2000a,c; S\'anchez \& Parravano 1999). There are also analytical models of
the IMF that include many physical processes involved with star formation
(e.g., Silk 1995; Nakano, Hasegawa, \& Norman 1995; Adams \& Fatuzzo 1996;
Myers 2000a).

Several reviews of the IMF are in the conference series {\it The Stellar
Initial Mass Function}, edited by G. Gilmore and D. Howell (Cambridge
University Press, 1998). Cayrel (1990) has an earlier review the theory of
the IMF, and Leitherer (1999) has a recent summary of IMF observations in
starburst galaxies. Other reviews are in Elmegreen (1998; 1999c; 2000d).

This paper discusses the basic classes of IMF models and various
observational constraints for them. These constraints suggest that the
IMF is approximately uniform in diverse environments, the star formation
process is relatively rapid and possibly independent for each star,
even in clusters, and that stars sample the IMF randomly in each cloud.
The recent observation of IMF-like mass functions in pre-stellar condensations
suggests further that the IMF may be decided in the gaseous phase, and
not be part of the collapse process during which this gas gets converted
into stars. These observations limit the physical processes that can be
involved in determining the IMF.

\section{Review of Theories}

\subsection{Four Types of Models}

Theories of the IMF may be categorized into several distinct types. Some
begin with a model for the formation of a single star and then vary the
parameters to get a range of stellar masses. These variations can be
random, as in the models by Larson (1973), Elmegreen \& Mathieu (1983),
Zinnecker (1984), and Adams \& Fatuzzo (1996), or they can depend on time
(Silk 1977) or position (Padoan et al. 1997). The statistical aspect
of the random models is realistic, but the physical models for star
formation in these pictures are often simplistic.

Other theories begin with a model for clustered star formation and
consider protostar interactions (Silk \& Takahashi 1979; Bastien 1981;
Yoshii \& Saio 1985; Lejeune \& Bastien 1986; Allen \& Bastien 1996;
Price \& Podsiadlowski 1995; Murray \& Lin 1996) or parameter variations
for single stars induced by multiple winds (Silk 1995). The good thing
about these models is that most stars are born in clusters, but the
interaction theories are often simplified or unphysical. The older
theories, for example, assumed the clumps would stick after a collision,
but if unbound clumps are transient density structures in a compressibly
turbulent gas, or if the clumps are gravitationally bound but move at
the larger virial speed of a whole cloud core, then mutual collisions
should destroy them. Most stars seem to form too quickly and to have
relative speeds that are too low (Belloche et al. 2000, this conference)
to interact like this anyway.

A third class of models considers clustered star formation with
competitive accretion (Larson 1978; Tohline 1980; Bonnell et al. 1997;
Myers 2000a). In these models, protostars move around in a uniform sea of
gas, which is the cloud core, and accrete it as they go. Accretion is a
well-accepted process of star formation, but the idea of a near-uniform
reservoir of gas for shared accretion contradicts the observation that
individual or binary stars form in very dense and separate clumps (i.e.,
in the pre-stellar condensations: Motte, Andr\'e, \& Neri 1998; Testi \& Sargent
1998; Belloche et al. 2000; Tachihara et al. 2000). The interclump
medium is at a much lower density than the clumps anyway, and would
not be a good reservoir for clump growth if the only relative speed is
from clump motions.  Supersonic interclump motions from compressible
turbulence should change clump masses much more rapidly than clump
motions.  In any case, most stars probably just get some variable fraction
of the mass of the dense clump in which they are born, perhaps between
3\% and 30\%, and that is all there is to accretion. {\it The accretion
rate does not matter if the reservoir of gas is limited to the clump},
and the accretion is not competitive if the clumps have a small volume
filling factor, like a few percent, which is observed to be the case
in the studies by Motte et al. (1998) and  Testi \& Sargent
(1998). In addition, {\it the range for the mass fraction of the clump
that gets into a star is probably much too small to explain the whole IMF:
most of the IMF has to come from the range of protostellar clump masses.}

This leads us to a fourth class of theories, in which much of the IMF
is attributed to cloud or clump mass functions. In the older versions
of this theory, the star mass scaled with some power of the clump mass,
and the clump mass spectrum observed in CO surveys was used to give the
star mass spectrum (Zinnecker 1989; Nakano, Hasegawa, \& Norman 1995). The
advantage of this model is that the structure of clouds can be observed
directly. However, the interpretation of this structure in terms of mass
functions for gas is very difficult (see Sect. \ref{sect:ms} below).
A second problem is that the origin of the clumpy structure in the gas
is not well understood, so this solution for the IMF merely replaces one
unknown with another. Moreover, if cloud structure dominates the IMF,
then the star formation process may hardly affect it. That leads to the
disappointing possibility that a correct and successful theory of the IMF
slope will not, in the end, reveal much about star formation. It will tell
us primarily about clump formation and the origin of cloud structure.
A new class of theories based only on sampling cloud structure (Sect.
\ref{sect:model}) has this aspect.

\subsection{A Characteristic Mass for Star Formation}

The IMF has essentially three measurable parameters: the average slope of
a power law part at intermediate to high mass, a different average slope
at low mass, down to and below the brown dwarf mass, and an intermediate
slope and characteristic mass that separates these two regimes. The
intermediate and high mass slope has received the most attention by
theoreticians, and models discussed above can usually reproduce it with
enough free parameters. The slope at low mass is only recently known and
only one theory so far has been proposed to explain it (Elmegreen 2000a;
see also the discussion in Myers 2000b).

The characteristic mass of a star at the border between these two
power law parts, which is $\sim0.3-1$ M$_\odot$, would seem to be
relatively easy to explain. There are very few characteristic masses in
the interstellar medium, which mostly shows power-law behavior over
a wide range of scales. Indeed, there are only two recent models that
get this mass. One suggests that cloud pieces collapse to much smaller
objects first, and that accretion continues until the star begins to
limit its own mass. This self-limitation defines a characteristic mass,
and models give it about the right value following Deuterium burning
in the protostar (Larson 1982; Shu et al. 1987; Nakano et al. 1995;
Adams \& Fatuzzo 1996). Note that in these models, the characteristic
mass is not actually the mass at the threshold for Deuterium burning,
which is much less, 0.018 M$_\odot$ (D'Antona \& Mazzitelli 1994). An
advantage of this model is that the characteristic mass is determined
partly by the star itself and is not overly sensitive to properties of
the surrounding cloud. Of course, higher pressure environments should
have higher density cloud cores, and the higher pressures inside of them
could resist the budding wind more at an early stage, thereby allowing
more massive stars to form. Higher temperature clouds could accrete
mass at a greater rate, holding back the wind too. But self-limitation
should dominate these cloud processes if pre-stellar winds are strong,
and this implies that the IMF might have a universal quality, independent
of environment.

The other theory for a characteristic mass in the IMF is that it is
linked to the smallest possible self-gravitating mass in a cloud (Larson
1992; Elmegreen 1997, 1999a, 2000a,c). Clouds consist of structures
on a wide range of masses, from far below the smallest star to far
above the largest star. Diffuse molecular clouds are known to contain
structures down to $10^{-4}$ M$_\odot$ (Heithausen et al. 1998), and
self-gravitating clouds may have structures down to $10^{-3}$ M$_\odot$
(Langer et al. 1995). Some of the clumps seen with molecular emission
lines could be the result of velocity crowding (Pichardo et al. 2000)
and not be real physical objects, but the tiny clumps in 100 $\mu$
IRAS maps of diffuse clouds could not have this origin because IRAS
integrates over the whole line of sight.  The smallest cloud clumps
even in molecular clouds are probably not self-gravitating (Bertoldi \&
McKee 1992; Falgarone, Puget, \& P\'erault 1992), in which case they are
not likely to form protostars. To become unstable, these smallest clumps
have to be compressed by pressures that are much larger than anything
likely to occur in a local star-forming region. Because of this, {\it
the origin of the stars that form from clumps less massive than the
smallest self-gravitating piece in a cloud should differ from that of
the more massive stars.}

The smallest self-gravitating mass in a cloud is about the thermal
Jeans mass or Bonner-Ebert mass, which depends primarily on the
thermal temperature and the total pressure: $M_J\sim0.3\left(T/10\;{\rm
K}\right)^2\left(P/10^6\;{\rm km \; cm}^{-3}\right)^{-1/2}$. Cloud pieces
much smaller than this should not begin the star formation process unless
an excursion in $P$ causes $M_J$ to decrease locally. For typical cloud
parameters, using the total turbulent and magnetic pressure for $P$ in
this expression ($P_{\rm total}\sim10^6$ K cm$^{-3}$), the thermal Jeans
mass is at the inflection point of the IMF. The sharp rise in star counts
with decreasing star mass, going from massive to intermediate mass stars,
ends at this point. It is followed by an IMF that is flat or slightly
falling toward lower masses on a log-log plot (e.g., Luhman \& Rieke 1998,
1999; Muench, Lada \& Lada 2000; Hillenbrand \& Carpenter 2000; Kaas \&
Bontemps 2000).

Sometimes a mathematical formula cannot be used to connect the parameters
in a cloud, such as density, pressure and temperature, with a final
stellar mass that forms after a long chain of events. Cloud conditions are
turbulent and chaotic, perhaps like atmospheric clouds on the Earth. Then
the approximate initial conditions for star formation may not completely
specify the outcome of cloud evolution.  It might be no more possible
to predict the final stellar mass from initial cloud conditions than to
predict rainfall amounts in New York from the humidity and temperature of
the same air when it is in Chicago two days earlier. The divergence in
outcomes of two infinitesimally close initial conditions in a turbulent
medium is well known. {\it This divergence problem should be particularly
troublesome for theories of the IMF that give a functional form to a
star's mass based on cloud parameters and then vary the parameters. }

For a characteristic mass inside the cloud, the problem with turbulent
chaos should not be too bad. The thermal Jeans mass or Bonner-Ebert mass
should be accurate enough for most purposes if it is applied to a cloud
structure that currently has the assumed temperature and pressure. The
result gives a local minimum for a self-gravitating clump mass, and
it provides a starting point for a theory on the IMF inflection. The
individual stellar masses are conceptually far removed from this cloud
mass, considering the additional processes that must come into play
before a star forms. So the connection between $M_J$ and any star mass,
even at the inflection point, remains vague.  An easy way out is to
suppose that $M_J$ has meaning only for cloud processes, and that once a
self-gravitating cloud piece forms, all of the additional complications
regarding collapse and star formation produce only a modest range for
the conversion factor from the mass of that cloud piece into a star
mass. Then the mass of the piece would still determine the {\it average}
mass of the star that forms in it, perhaps to within a factor of $\sim3$.
{\it The distribution function for star mass inside a cloud piece of
a given mass should be viewed as another property of star formation,
like the IMF, but not necessarily equal to the final IMF, which also
contains information about the origin and evolution of the cloud pieces}
(Elmegreen 2000a).

Another problem with the $M_J$ limit is that stars form at much lower
masses too, so $M_J$ is not a hard barrier to star formation. This may
not mean that smaller cloud pieces turn independently into stars. The
smaller stars may arise from fragmentation inside $M_J$ pieces (more on
this below).

Magnetic forces and turbulent motions should increase the self-gravitating
mass above $M_J$, and a decrease is possible at the center of a converging
flow (Hunter \& Fleck 1982). If the total pressure used in the expression
for $M_J$ includes these turbulent motions, then the result should still
be a useful lower limit to self-gravitating cloud structure.

A few years ago there was a third model for a characteristic mass in star
formation, the mass of an optically thick core at the limit of strong
self-gravity (Rees 1976; Yoshii \& Saio 1985).  This mass is very small,
$\sim10^{-3}$ M$_\odot$, so models that used it had to rely on coalescence
or continued accretion to achieve final stellar masses.  In that case,
the final characteristic mass will be more closely related to the other
characteristic masses given above than to the opacity-limited mass.

\section{Observational Constraints}
\subsection{IMF Uniformity}
\label{sect:unif}

The IMF is approximately uniform in space and time, as illustrated by its
similar form in galactic clusters (Scalo 1998), OB associations (Massey
1998), old globular clusters (de Marchi \& Paresce 1997; Chabrier \&
Mera 1997; Pulone, de Marchi, \& Paresce 1999; De Marchi, Paresce,
\& Pulone 2000; Paresce \& De Marchi, 2000), halo stars (Nissen et
al. 1994), and bulge stars (Holtzman et al. 1998). The near-uniformity
of the IMF has also been demonstrated by a similar ratio of iron to oxygen
abundance, which is a measure of the ratio of low-mass to high-mass stars,
in elliptical galaxies (Wyse 1998), the intracluster medium (Renzini,
et al. 1993; Wyse 1997, 1998; but see Loewenstein \& Mushotzky 1996),
and QSO Ly$\alpha$ absorption systems (Lu et al. 1996; Wyse 1998). The
IMF is also independent of metallicity (Freedman 1985; Massey, Johnson \&
DeGioia-Eastwood 1995a).

\begin{figure}
\vspace{3.2in}
\includegraphics{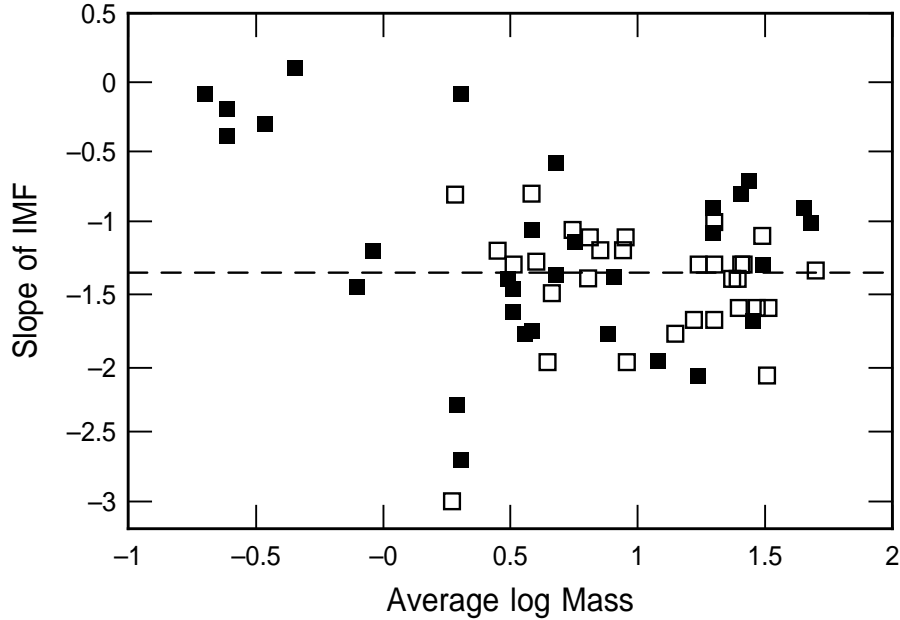}
\caption{IMF slopes in different clusters as a function
of the average log mass, in M$_\odot$, from Scalo (1998).
The Salpeter value of $-1.35$ is shown by a dashed line.  Solid squares
are for clusters in the Milky Way, and open squares
are for the Large Magellanic Cloud.}
\label{fig:scalo}
\end{figure}

The IMF is usually a power law at intermediate to large mass, with a
slope between $\sim-1$ and $\sim-1.5$ on a log-log plot. At low mass
the IMF flattens to a slope of $\sim0.5$ to $-0.5$. A compilation of
IMF slopes from a variety of clusters was made by Scalo (1998) and
is reproduced here in Figure \ref{fig:scalo}. Each point represents an
average mass for the observed part of the cluster versus the slope of the
IMF centered on this average mass. Deviations of $\pm 0.5$ from point to
point around the Salpeter slope of $\sim-1.35$ are present. These could
be statistical fluctuations, considering the small number of stars that
are usually measured.

\begin{figure} \vspace{3.2in} 
\includegraphics{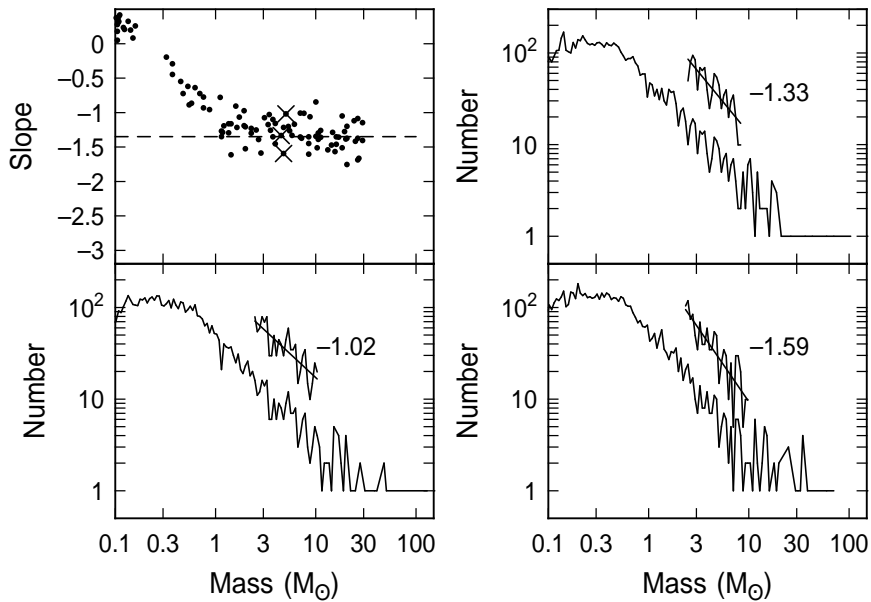} 
\caption{(top left) IMF slopes in 100 models plotted as
a function of the average logarithm of the mass. Each IMF slope is
fit using 200 stars.  The three values indicated by crosses have their
complete IMFs shown in the other panels, with the fitted portions of these
IMFs indicated by the offset lines.} \label{fig:scatter} \end{figure}

For example, Figure \ref{fig:scatter} shows a similar plot that is
based on a random sampling model for cloud pieces in a fractal cloud,
as described below (Elmegreen 1999a). Each point represents the IMF
from the brightest 200 members of a model cluster, and there are 100
clusters of various total masses contributing to the plot. More massive
clusters have bigger most massive stars, and so populate the right hand
part of this diagram, as in the real observations. The scatter in the
models is about the same as the scatter in the observations. The other
three panels in Figure \ref{fig:scatter} show the model IMFs that are
represented by the crosses in the top-left panel. Both the steep and the
shallow IMFs look statistically significant in these plots, but they are
only random variations around the average slope, which is the Salpeter
value in the model.

{\it Steeper IMF slopes are sometimes found in lower density regions,}
including the local field (Garmany, Conti, \& Chiosi 1982; Humphreys \&
McElroy 1984; Scalo 1986; Blaha \& Humphreys 1989; Basu \& Rana 1992;
Kroupa, Tout, \& Gilmore 1993; Parker et al. 1998) and some low-density
parts of clusters (J.K. Hill et al. 1994; R.S. Hill et al. 1995;
Ali \& DePoy 1995). This difference may result from mass segregation
in clusters, or from differential drift of the low and high-mass stars
away from their points of origin.  For example, the IMF is known to be
shallower in dense cluster cores than along the cluster periphery (Sagar,
et al. 1986; Jones \& Walker 1988; Sagar \& Bhatt 1989; Sagar \& Richtler
1991; Pandey, Mahra, \& Sagar 1992; Vazquez et al. 1996; Fischer et
al. 1998; Kontizas et al. 1998; Hillenbrand \& Hartmann 1998). There is
no satisfactory explanation for this segregation of high-mass stars to the
center; normal two-body interactions do not seem to work fast enough to do
this (Subramaniam, Sagar, \& Bhatt 1993; Fischer et al. 1998; Bonnell \&
Davies 1998; although see Giersz \& Heggie 1996). 
Competitive accretion might do it because gas accretion provides
a drag on stellar orbits, and this drag is larger for stars that
accrete more, causing them to migrate closer to the center of the cluster
(Bonnell et al. 1997). However,  
if the accreting gas has random speeds like the protostars
and is not perfectly static, then the orbital energy 
densities of the protostars and the accreting gas 
will be the same. In this case, there should be no 
additional mass segregation with
competitive accretion over and above the root-N decrease in 
center-of-mass energy that arises simply from the presence of
more sub-clumps in more massice stars. That is, random sampling of
turbulent gas by any mechanism will 
give a slight mass segregation all by itself
because the average momentum per unit mass of
a cloud piece relative to the cloud center is smaller if
that cloud piece contains a larger number of random gas elements, 
i.e., if it is more massive. 
Differential drift can also
give the field stars a steep IMF, because low-mass stars live longer
and drift further from their formation sites than high-mass stars.

A very steep IMF slope of $\sim-4$ (compared to the Salpeter slope of
$-1.35$) was found in extreme field regions far from known clusters and
associations in the LMC and local field (Massey et al. 1995b). There
are simple explanations for this extreme steepening based on cloud
destruction reviewed below, but the observations need to be confirmed
before the models can be carried much further.

Perhaps the most interesting of the proposed deviations from a universal
IMF is a shift toward higher mass stars in starburst regions. This shift
could take the form of a more shallow IMF in the intermediate to high-mass
range, or it could be a parallel shift of the whole IMF toward higher mass
without a change in the shape. The main reason the IMF looks different
in starbursts is the large ratio of luminous to dynamical mass (Rieke et
al. 1980, 1993; Kronberg, Biermann, \& Schwab 1985; Wright et al. 1988).
Confirmations of this proposal have come from galactic evolution models
(Doane \& Matthews 1993), spectroscopic line ratios (Doyon, Joseph, \&
Wright 1994; Smith et al. 1995), and infrared excesses (Smith, Herter
\& Haynes 1998). However, Devereux (1989) and Satyapal et al. (1995,
1997) lowered the extinction correction for M82, and this makes the IMF
there more normal. Other recent studies involving evolutionary models
(Schaerer 1996), multiwavelength spectroscopy and broad-band infrared
photometry (Calzetti 1997), and emission line spectroscopy (Stasi\'nska
\& Leitherer 1996) give normal IMFs too. A general problem with large
IMF shifts in starburst galaxies is that they should produce unobserved
red populations of stars after the turnoff age reaches the stellar
lifetime at the truncation mass (Charlot et al. 1993). The ISM should
also develop too high an oxygen abundance compared to iron because of
the overabundance of high-mass stars (Wang \& Silk 1993).

Other indirect observations have led to theoretical predictions of
IMF changes. Fabian (1994) proposed that the IMF is biased toward
low mass stars in galaxy cluster cooling flows, and Elmegreen (1999b)
proposed a similar bias in ultracold molecular gas, both because of a
low thermal Jeans mass that is expected in these regions. In the first
case, the high pressure lowers the Jeans mass and in the second case the
low temperature does. Larson (1998) and Bromm, Coppi, \& Larson (1999)
proposed that the IMF shifted toward higher mass in the early Universe,
and that this would help explain the G-dwarf problem
(i.e., the presence of metals in the oldest Galactic halo stars),
the high temperature
and high metal abundance of intracluster gas, and the large luminosities
of young elliptical galaxies.

A theory of the IMF should be flexible enough to accommodate changes
in different environments. What actually changes is not yet known,
however. Of the three aspects of the IMF mentioned above, i.e., the
high and low-mass slopes and the characteristic mass, the latter alone
may cause most of the observed changes. The characteristic mass for a
star depends on the physical properties of star formation, and perhaps
the upper and lower mass limits depend on these properties too, so
we might expect differences in extreme environments that could shift
the characteristic mass either up or down without changing the slopes
much. In normal regions, however, the general similarity of the IMF from
place to place and over time suggests that the basic process by which
the gas partitions itself into stars is somewhat universal.

\subsection{Independent Formation of Stars even in Dense Clusters}

The observed correlations between cluster density and maximum star mass
(Testi, Palla \& Natta 1999), and between cluster mass and maximum star
mass (Larson 1982), give the appearance of collective effects during
star formation, as if stars influence each other, perhaps by radiative
effects or coagulation in dense cluster cores. However, the actual
observations are no more correlated than what is expected from random
sampling (Elmegreen 1983, 1997, 2000c; Schroeder \& Comins 1988; Massey
\& Hunter 1998; Selman et al. 1999; Bonnell \& Clarke 1999), so there
is no real evidence that any young stars care about their neighbors,
except possibly for binary stars.

A purely statistical IMF in a Salpeter power law leads to a correlation
between cluster mass (in the power-law range of the IMF) and maximum
star mass of the form (Elmegreen 2000c): 
\begin{equation}M_{cluster}\sim
3\times10^3 \left({{M_{max}}\over{100\;{\rm M}_\odot}}\right)^{1.35}
{{\rm M}_\odot}.
\label{eq:mmax}
\end{equation} 
This means
that bigger clusters form bigger most-massive stars simply because they
sample further out into the tail of the IMF. The correlation between
cluster density and maximum star mass found by Testi, Palla \& Natta
(1999) could, in principle, be the result of IMF sampling, because most
of their clusters have about the same radius, 0.2 pc, so cluster density
is proportional to mass (see also Bonnell \& Clarke 1999).  There is
no explanation for this constant radius, and it is a surprising result
given that CO {\it cloud} radii generally correlate with the inverse
square roots of their masses (Larson 1981).

The IMF itself, aside from the maximum star mass, appears to be relatively
independent of cluster density over a factor of ~200 in density (Hunter,
et al. 1997; Massey \& Hunter 1998; Luhman \& Rieke 1998).

\begin{figure}
\vspace{2.6in}
\includegraphics{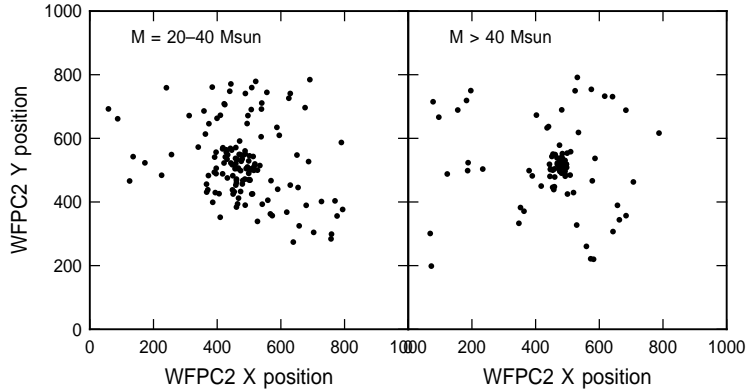}
\caption{Positions of massive stars in the R136 field of 30 Dor
taken from the WFPC2 camera. The positions are given in units
of camera pixels, where the pixel size is 0.04555''. The data
were kindly provided by D. Hunter from the publication
Massey \& Hunter 1998. }
\label{fig:30dor}
\end{figure}

Another reason one might get the impression that star neighbors influence
final star mass is that the Orion cluster has its most massive star in
the center, in a very crowded field of other stars. Perhaps this star
got to be so massive because of some influence from all of the other
stars. However, stars as massive as the O5 star in Orion appear all over
the 30 Doradus cluster in the LMC and not just in the center, as shown in
Figure \ref{fig:30dor}. These peripheral O-type stars may have been born
in the cluster core and drifted out into this surrounding region over
time, but until this is demonstrated by a comparison of stellar ages, we
cannot conclude that all massive stars are born in dense cluster centers.
There is an excess concentration of massive stars in the center of this
cluster, but no evidence yet that these stars formed only in the dense
cluster core as a result of, say, coalescence or enhanced accretion.

There is some evidence that collisions might be important in dense
clusters, but no evidence yet that the IMF will be affected by this.
The collision time between pre-stellar condensations in the Ophiuchus core
region seems to be fairly short compared to the total star formation time.
The angular filling factor for Ophiuchus condensations in cores A and
B of the survey by Motte et al. (1998) is about $10$\%. This means
that the number of half-orbits that a condensation has to undergo
before colliding physically with another condensation is the inverse
of this filling factor divided by the gravitational focussing factor,
$1+2GM/\left(Rv^2\right)\sim7$, for protostellar mass $M\sim0.3$
M$_\odot$, radius $R\sim0.01$ pc, and mutual speed $v\sim0.2$ km
s$^{-1}$. These numbers imply that pre-stellar condensations should
collide after only about one crossing time, which is fast enough for
coalescence to be important. However, the density of a pre-stellar
condensation is enormous, $\sim3\times10^7$ H$_2$ cm$^{-3}$, so the
internal dynamical time of each one is only $\sim 0.01$ My, which is
one-tenth of their mutual collision time. Thus the observed pre-stellar
condensations should collapse significantly before they interact,
leaving only rotationally-supported disks at the presently observed size
of $\sim0.01$ pc.  These disks will probably interact in the manner
discussed above, but the stars inside of them should be too small to
coalesce.  {\it Thus the IMF will probably not be affected by direct
stellar coalescence, even in a dense region like the core of Ophiuchus.}
The nature of protostellar disks could be affected by cluster density,
however.

If the star-formation time in a dense core were much longer than a
crossing time, say $\sim100$ crossing times, then stellar coalescence
might be more important. But there is no evidence for such prolonged star
formation.  On the contrary, the available evidence seems to suggest that
star formation is very rapid.  This comes from direct observations of
young-star ages in embedded clusters, from the high fraction of clouds
with young stars and the rapid dispersal of gas after star formation,
and from the observation of hierarchical structure in young star fields,
which indicates very little star-star mixing (Elmegreen 2000b). There is
also a correlation between the duration of star formation and the size
of the region (Elmegreen \& Efremov 1996; Efremov \& Elmegreen 1998),
which is essentially the same as that between the turbulent crossing
time for gas and the region size. This correlation indicates that the
duration of star formation is always about the turbulent crossing time
for a wide range of scales.

\subsection{The IMF as a Statistical Ensemble}
\label{statemsemb}

We have discussed above how the most massive star in a cluster appears to
come from IMF sampling only; more massive clusters produce more massive
stars primarily for this reason.  If this is really a statistical
effect, then sometimes a low-mass cloud will produce a high-mass star
-- a result that seems counterintuitive. However, the interstellar gas
is often structured in a self-similar and hierarchical way, so massive
clouds contain lower-mass sub-clouds, and these contain even lower mass
sub-sub-clouds, and so on.  When viewed from a great distance, a whole
young star field may appear to be forming in a giant cloud complex,
but when viewed from nearby, stars of various masses should be seen
forming in sub-parts of the cloud having various smaller masses, in an
overall random fashion. In this sense, some of the smaller subpieces,
perhaps the size of the Taurus clouds, should be forming massive stars,
unlike the Taurus region itself. But these subpieces are parts of larger
clouds too. Which of the various cloud masses and sub-cloud masses should
be counted in this statistical interpretation?

The main point is that as far as we can tell, a star of any mass
seems to be able to form in a cloud of any mass, provided there is
enough material to make the star. This means that the summed IMF from
10 clouds with only $10^4$ M$_\odot$ each should be the same as the
IMF from a single cloud with $10^5$ M$_\odot$. The problem here is not
with the star formation part, but with defining cloud mass in a fractal
distribution of density. Indeed, the average IMF in a whole galaxy is
indistinguishable from the IMF in a single cluster. Whole-galaxy IMFs come
from the color-magnitude diagrams of nearby dwarfs (Greggio et al.  1993;
Marconi et al. 1995; Holtzman et al. 1997; Grillmair et al. 1998),
from the equivalent widths of H$\alpha$ in galaxies (Kennicutt, Tamblyn
\& Congdon 1994; Bresolin \& Kennicutt 1997), and from the iron/oxygen
ratio, as discussed above. This equivalence means that the summed IMFs
from many small clouds and clusters is similar in form to the IMF from
any one of them.  This can only occur if stars of all masses have equal
probabilities of forming in clouds of all masses (Elmegreen 2000d).

An observation like this rules out many IMF models in which stars
influence each other. For example, if cloud destruction by massive stars
halts star formation, and if larger clouds need more massive stars for
their destruction (because of their greater self-binding), then star
formation will proceed in a low-mass cloud, building up more and more
massive stars until a particular star mass is reached, and then stop
without making any more massive stars.  Elsewhere in a more massive cloud,
the IMF will proceed further, making stars of the same mass range as the
lower mass cloud, but also higher mass stars before cloud destruction.
In a scenario like this,  the summed IMF will be steeper than the IMF
in either cluster because there are a lot of low-mass clouds that will
add only to the low-mass part of the total IMF.

Also in a model in which star mass increases as more and more stars form,
perhaps as a result of enhanced heating from all the stars, we would
have a similar result that only massive clusters can produce high mass
stars. But then the sum of the IMFs from low-mass clusters and high mass
clusters would be steeper than either one, contradicting the observation
that these IMFs are the same.

The steepening of the IMF in the extreme field regions studied by Massey,
et al. (1995b) could be an example where the summed IMF is in fact
steeper than each component. What differs about these regions is their
extremely low pressures.   Perhaps clouds are more easily disrupted by
star formation at such low pressures: the observed IMF slope in these
field regions is consistent with cloud destruction by ionization because
the destructive power of OB stars increases with star mass in the correct
fashion (Elmegreen 1999a). The contribution of these few field stars to
the total IMF in a galaxy is small, however.

The IMF in the general field is also somewhat steeper than in clusters
(cf. Sect. \ref{sect:unif}). This is observed directly, before correcting
for mass segregation or differential drift.  Perhaps this implies there
is a self-limitation of stellar mass by other stars, but at the present
time, the field and galaxy-wide IMFs are not well enough known to give
any definitive insight into this possibility.

\section{On the Meaning of Mass Spectra in a Fractal Medium}
\label{sect:ms}

Clump-finding algorithms do not consider nested structures like the
hierarchical fractals of real interstellar clouds. Algorithms like
these find only those structures within a factor of $\sim3-10$ of the
telescope beam size (Verschuur 1993; Elmegreen \& Falgarone 1996). Larger
structures are resolved out and ignored, while smaller structures are
not observable. These algorithms give a false impression of what cloud
structure is like. We know from power spectra of emission maps of clouds
that there is structure on a much wider range of scales than the clump
sizes that come from contour drawings of the same region (Stutzki,
et al. 1998). Yet the
{\it mass} spectra made from these clump-finding algorithms look
reasonable: they often span a factor of $\sim100$ in mass and define
a nice power law. However, even this mass power law gives a false
impression. The mass scales with about the 2.3 power of the size of a
region (this power is presumably the fractal dimension; Elmegreen \&
Falgarone 1996; Heithausen et al. 1998), and so a size spectrum with
only a narrow range around the telescope beam size can turn into a mass
spectrum with a fairly wide range. Yet, the mass spectrum is just as
biased by the beam as the size spectrum because they both come from the
same clump identifications that ignore the scale-free fractal structure.

In hierarchical clouds, the mass spectrum of structures of all types is
$n(M)$ $d\log M $ $\propto$ $M^{-1}d\log M$ (Fleck 1996). This spectrum
has equal mass in equal logarithmic intervals of mass. Hierarchies give
this because the levels in the hierarchy are logarithmically spaced. For
example, one big piece divides into 4 smaller pieces, and each of these
divides into 4 more, giving 16 total, and then 64, etc.. The same mass
is counted again and again in each level.  Such perfect hierarchical
structure may not apply to real interstellar clouds, but cloud structure
is hierarchical (Scalo 1985), and the ideal case is conceptually useful.

Multiple counting of mass for hierarchical clouds is bad if the sum of the
clump masses must equal the cloud mass, but it is good if all we want is
a probability distribution function for clump mass. If structures at any
level in an ideal hierarchy are randomly chosen, then the probability
distribution function for mass $M$ is $M^{-1}d\log M\equiv M^{-2}dM$.
Mass spectra of clumps found by contour-drawing methods, which tend to
find clumps only near the beam size as discussed above, differ from
mass spectra of randomly sampled pieces in a self-similar hierarchy
of structures, presumably because of the different ways the structures
are defined.

An example of a clump mass spectrum that multiply counts pieces and also
gets a slope close to the expected value of $-2$ was given by Heithausen,
et al. (1998). They studied a region in the Polaris Spur using three
telescopes with different beam sizes. One of the clumps found with the
largest beam was observed again with a smaller beam, and one of the
small clumps observed with this smaller beam was observed a third time
with an even smaller beam. All of the structures were put on the same
mass spectrum and the resulting slope was $-1.85$, much steeper than
the usual $\sim-1.5$ for cloud pieces.

Star clusters and OB associations are the result of star formation in
locally dense regions of interstellar gas. These dense regions cluster
together on a wide range of scales because of the fractal nature of the
gas, and this causes the stars to cluster in the same way (see review
in Elmegreen et al. 2000e). Clusters and OB associations sample from
all over the hierarchy of gas structures, and because of this, end up
with an approximately $M^{-1}d\log M$ spectrum (Kennicutt, Edgar, \&
Hodge 1989; Battinelli et al. 1994; Comeron \& Torra 1996; Elmegreen \&
Efremov 1997; Feinstein 1997; McKee \& Williams 1997; Oey \& Clarke 1998).
Such a hierarchical clustering of stars offers a good example of how mass
structures can be counted without the telescope beam bias that is present
in contouring algorithms. Clusters randomly sample the fractal hierarchy
of gas, and have the expected $\sim M^{-1} d\log M$ distribution.

\section{A Random Sampling Model for the Initial Mass Function}
\label{sect:model}

Stars also sample mass from the hierarchy of gas structures, and would
presumably have an $M^{-1}d\log M$ spectrum like the clusters except
the stars compete with each other for gas. Once a star forms in part
of a hierarchy, the mass remaining in higher levels of the hierarchy
no longer has that gas available to make another star. So any other
star that forms later at a higher level, which would normally have more
mass than the first star because it includes many lower level clumps,
would find itself short one of these clumps (because that clump is
now an independent star), and as a result, this next star would have a
slightly lower mass than its initial share.  This systematic lowering of
the masses of all late-forming stars will steepen the IMF a little if
the late-forming stars are slightly more massive than the first stars,
on average. It does not violate the premise that stars of all masses form
in clouds of all masses (Sect. \ref{statemsemb}), because the steepening
process occurs {\it during} the conversion of a hierarchical cloud into
stars, whereas the comparison between galactic and cluster IMFs discussed
above is done after the cluster formation process is mostly over.

Clusters complete with each other for mass too, but the mass of a final
cluster is taken to be the sum of the masses of its parts. That is,
the cluster mass is the whole mass in a certain part of the hierarchy of
structures. It includes all of the lower mass sub-clusters in its total,
unlike the late-forming stars. So, the cluster mass is the whole
hierarchy mass, while the star mass is the hierarchy mass minus the masses
of the other stars that formed first in lower parts of the hierarchy.

This description of star formation is made for an idealized static model,
whereas clumps come and go on a local dynamical time, and the conversion
of gas into stars operates on a dynamical time too. However, the result
is the same whether we speak of a static model or a dynamical model. In
a static model, there are more small pieces than large because there is
more room for the small pieces, whereas in a dynamical model, there are
more small piece than large because the formation rate of the small pieces
is much larger than the formation rate of the large pieces that contain
them. The difference between the two models reflects one's point of view,
not the mass functions for cloud structure.  Thus, numerical simulations
in which supersonic turbulence continuously generates fractal structure
should give the same mass spectra for structures at any one time as they
do for all of the structures that form over time.

The final result of gas competition for star formation depends on
the relative birth order of the stars. If all stars form at the
local dynamical rate, $(G\rho)^{1/2}$, then denser regions tend to
form stars first. This would be the case for gravitational processes,
or for kinematical processes in virialized regions including magnetic
diffusion, clump collisions, and turbulence compression. In a fractal,
hierarchical cloud, the density depends on the mass: $\rho \propto
M^{1-3/D}$. For $D<3$, as observed in interstellar clouds, the low-mass
regions are denser and proceed toward star formation faster than the
high-mass regions that contain them. For this reason, an IMF based on
random sampling in fractal clouds is steeper than $M^{-1} d \log M$;
i.e., low-mass stars are more competitive for the available gas than
high-mass stars.  Numerical simulations of this competitive process give
the Salpeter IMF, i.e., $\sim M^{-1.35} d\log M$ (Elmegreen 1997, 1999a).

\begin{figure}
\vspace{4.4in}
\includegraphics{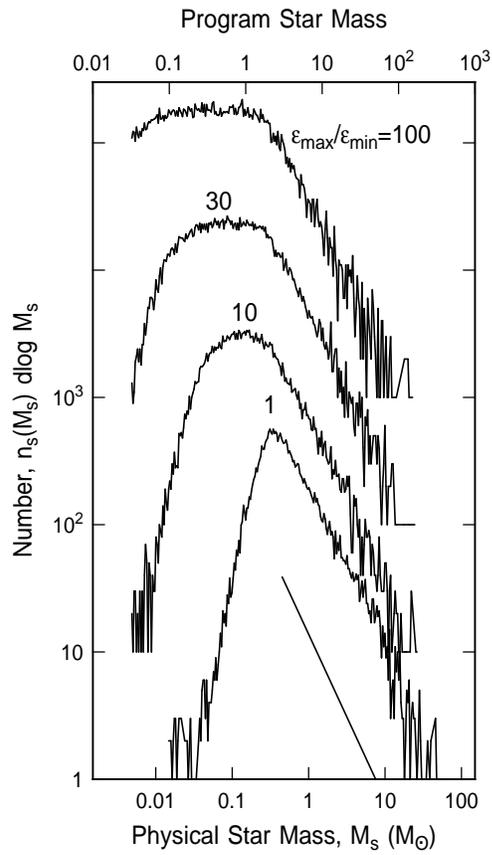}
\caption{IMF model with random sampling from a
hierarchical cloud for four values of the mass range ${\cal R}$,
defined to be the ratio of the maximum fraction to the minimum 
fraction of the clump mass that goes into a star. The straight line
has a slope of 1.3.}
\label{fig:flat}
\end{figure}

The scaling relations for molecular clouds suggest that the
intermediate-to-high mass part of the IMF comes from a factor of $\sim7$
in length scale. This is because $M\propto L^D$ for fractal dimension
$D\sim2.3$. This power law part of the IMF also comes from only a factor
of $\sim2.5$ in dynamical rate, because the dynamical time $\propto
M^{0.2}$ (Elmegreen 2000c). What this means is that essentially all
star formation happens on length and time scales that are close to
the threshold for self-gravity. This conclusion is consistent with the
overall rapid rate of star formation in molecular clouds, and it may
also have important consequences for the formation of high-mass stars,
discussed below.

The low-mass, approximately flat part of the IMF may be showing us some
type of fragmentation spectrum inside each clump (Elmegreen 2000a).  If
$\epsilon$ is defined to be the star-to-clump mass ratio and $P(\epsilon)$
is the distribution function for $\epsilon$, then the flat or slightly
falling (toward lower mass) part results if $P(\epsilon) d \log \epsilon
=$ constant or $\propto\epsilon^\kappa$ with $\kappa\sim0-0.5$ for all
star-forming clumps, including those whose stars contribute to the power
law part of the IMF.

Models are shown in Figure \ref{fig:flat} that were made by randomly
sampling mass structures in hierarchical clouds with a sampling rate
that is weighted by the local density, $\rho^{1/2}$ (from Elmegreen
2000a). Various ranges (${\cal R}=\epsilon_{max}/\epsilon_{min}$) for
$\epsilon$ are given, with $P(\epsilon) d \log \epsilon =$ constant,
and with $\epsilon_{max}$ and $\epsilon_{min}$ equal to the maximum and
minimum values of $\epsilon=M_{star}/M_{clump}$ in the distribution
function $P(\epsilon)$. If $\epsilon$ has a large range, then the
flat part of the IMF is long. The Salpeter slope always appears
at intermediate-to-high mass, with a turnover to a flat slope at
low mass.  This flat slope is entirely the result of our assumption
that $P(\epsilon)d\log \epsilon=$ constant. The important point is not
the value of $P$, for which there is no theory yet, but that the flat
or shallow part of the IMF shows up only below the thermal Jeans mass,
even though $P(\epsilon)$ has been applied to all clumps regardless of
mass. In this interpretation, the smallest stars or brown dwarfs come
from the smallest cloud pieces that are able to collapse into stars,
namely those with mass near $M_J$, and they arise for those few cases
where $\epsilon$ is near its minimum value.  The physical mechanism for
their formation is unknown.  They could form in the disks of low-mass
stars, for example.

At the high-mass end of the IMF, the power law Salpeter function drops
until there are very few massive stars in most clusters. According to
Massey \& Hunter (1998) and Selman et al. (1999), the high mass end
of the IMF in 30 Dor has not yet reached the limit of possible stellar
masses: the power law function continues to higher and higher mass until
there is only one most-massive star. Generally, bigger clusters continue
further, giving larger most-massive stars because they sample further
out in the IMF. This makes us wonder if there is a physical limit to the
mass of a star. Perhaps accretion can be optically thick, as in a disk,
and in this way get around the Eddington limit. Or perhaps supermassive
stars can form by coalescence (e.g., Zinnecker 1986), which is also an
optically thick form of accretion.  If the accretion is optically thin,
however, then the Eddington limit should influence the upper end of the
IMF (Norberg \& Maeder 2000). 

\begin{figure}
\vspace{3.3in}
\includegraphics{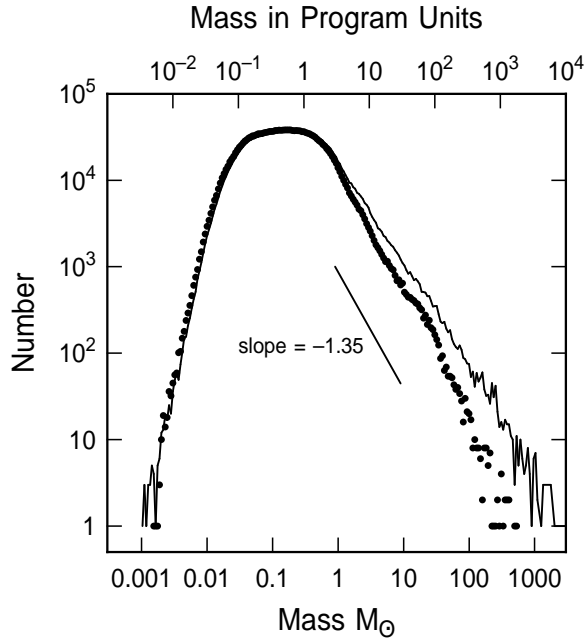}
\caption{Two IMF simulations are shown. The solid line is for a model
with no timing constraint in the choice of
clumps for star formation, while the
dotted line is with the timing constraint.
Both models contain $2\times10^6$ stars.}
\label{fig:biggest}
\end{figure}

There is a way to determine if stars have an upper mass limit: from
the statistics of star formation in a whole galaxy. Even though single
clusters do not have enough stars to sample out to a maximum stellar
mass of say, 1000 M$_\odot$, a whole galaxy does.  According to equation
\ref{eq:mmax}, a whole galaxy with $10^6$ M$_\odot$ of young stars should
produce a most-massive star with $\sim7000$ M$_\odot$.  Such stars are
not observed, even as pre-main-sequence objects, 
so the IMF cannot continue indefinitely to higher and
higher masses. Either there is a sharp cutoff just beyond the maximum
mass of $\sim120-150$ M$_\odot$ that is observed in 30 Doradus, or there
is a more gradual cutoff in most IMFs, with an excess of massive stars
compared to this in 30 Dor.
Note that internal disruption of a supermassive 
star by main-sequence or post-main-sequence
oscillations (Maeder 1985) will not prevent
a pre-main-sequence version of this star from forming; the
prevention of supermassive pre-main-sequence stars has to 
occur in the collapse or pre-collapse phase.

Aside from the Eddington limit, another
possible mass limit to a star comes from timing. As an absurd
illustration of this problem, one can ask why whole GMCs don't form
single stars. If the ISM is scale free, then they should. The answer
seems to be that other things happen first: lower mass stars take the
gas for themselves, and cloud disruption removes the remaining gas and
halts the whole process. These other things are important within only a
few dynamical times of the formation time of the lower mass stars. This
is the duration of star formation in most regions. However, $\sim2-3$
dynamical times includes the entire stellar mass range in the power law
part of the IMF, as discussed above. Thus stars much more massive than
$\sim100$ M$_\odot$ do not have time to form. Any clump of this mass that
is physically able to form a supermassive star has a large probability
of forming other, smaller stars first.

The IMF should begin to drop suddenly for stars that form in clumps
that are so massive that their dynamical times exceed $\sim3$ times
the dynamical time of an $M_J$ star. This is partly because turbulence
remixes the gas during this time, preventing the large turbulent clumps
from becoming uniform enough to form a single star, and it is partly
because star formation from lower mass stars, which tend to form first
statistically, uses up the gas and disrupts the cloud.

Figure \ref{fig:biggest} shows a simulation of this process in a
computer-generated IMF with $2\times10^6$ stars, made by randomly
sampling a hierarchical tree of gas structure with a $\rho^{1/2}$
sampling bias and no Eddington limit (from Elmegreen 2000c).  The dotted
line is made with an additional step in which a randomly chosen clump
of mass $M$ is allowed to turn into a star only with the probability
$\exp\left(-\tau\left[M\right]/\tau\left[M_J\right]\right)$. Clumps
that fail at this stage are sent back to the hierarchical gas tree
for further sampling, and are most likely to form several lower mass
stars instead. The solid line is for an identical simulation without
this additional step.  The dotted line falls fast enough at high mass
to avoid the formation of supermassive stars.  Such stars could still
exist if $M_J$ increases, because according to the theory, the high mass
cutoff scales with this characteristic mass.  Perhaps starburst galaxies
have supermassive stars.

\section{Conclusions}

The IMF has many properties suggestive of a connection to turbulence. It
is universal, stochastic, and power-law. Star formation also has dynamic
and structural properties similar to turbulence: the duration of star
formation in a cloudy structure is usually only a few crossing times,
regardless of scale, and starbirth positions are often fractal.

We have found that virtually any mechanism for star formation that
operates on a dynamical time scale in an initially fractal cloud will
produce the Salpeter IMF at intermediate to high mass. This IMF is a
consequence of random sampling from the various mass structures in the
cloud, and not a consequence of the star formation processes. Only the
mass at the lower limit of the power law part of the IMF, and the masses
at the upper and lower limits, should depend on the detailed physics of
star formation. The power law part could be determined in the gas phase.

The low-mass break in the power law part of the IMF might result
from a lack of self-gravity in tiny clumps. This break point
should be higher in regions with higher temperatures and/or lower
pressures. The flat-to-falling part of the IMF below the break may come
from fragmentation or other secondary processes inside the dense clumps
or disks where stars form, or from pressure fluctuations affecting the
Jeans mass. That is, low-mass stars and brown dwarfs should form along
with other more massive stars in the smallest cloud clumps. The lack
of supermassive stars in galaxies was considered to be a constraint on
the IMF. The IMF should fall off more rapidly than the Salpeter slope at
high mass because of timing limitations and the
Eddington limit. For the timing limit, clumps
whose internal dynamical timescales are longer than several times the
dynamical time of a clump with the characteristic mass, $M_J$, should
not proceed directly to form a single star. Smaller stars should form
in these clumps first, making the gas unavailable for the formation of
super-massive stars.

Observations of variations in the IMF in different environments would be a
good check on the theories. Extreme environments such as high-temperature
and high-pressure starburst regions, remote or low-pressure field
regions, highly compressed or triggered regions, 
ultracold gas, and high-pressure cluster cooling flows, might
shift the characteristic mass for star formation up or down, depending
on the thermal Jeans mass. If this is the only change that is observed
in these regions, then models where the power-law part of the IMF comes
from universal cloud geometry would be supported. If the power law slope
of the observed IMF changes with environment, however, and not just as a
result of mass segregation or selective cloud destruction discussed above,
then the IMF would not have the universal and scale-free aspect that is
expected from turbulence. Other IMF models that are more dependent on
the detailed processes of star formation would be required.


\begin{references}
\reference{} Adams, F.C., \& Fatuzzo, M. 1996, ApJ, 464, 256
\reference{} Ali, B., \& DePoy, D.L. 1995, AJ, 109, 709
\reference{} Allen, E.J., \& Bastien, P. 1996, ApJ, 467, 265
\reference{} Bastien, P. 1981,  A\&A, 93, 160
\reference{} Basu, S., \& Rana, N.C. 1992, ApJ, 393, 373
\reference{} Battinelli P., Brandimarti A. \& Capuzzo-Dolcetta R. 1994,
A\&AS, 104, 379
\reference{} Belloche, A., Andr\'e, P., \& Motte, F.  
2000, in From Darkness to Light: Origin and Early
Evolution of Young Stellar Clusters, eds. T. Montmerle \& P. Andr\'e, 
ASP Conf. Series, in press
\reference{} Bertoldi, F., \& McKee, C. 1992, ApJ, 395, 140
\reference{} Blaha, C., \& Humphreys, R.M. 1989, AJ, 98, 1598
\reference{} Bonnell, I.A., Bate, M.R., Clarke, C.J., \& Pringle, J. E.
1997, MNRAS, 285, 201
\reference{} Bonnell, I.A., \& Davies, M.B. 1998, MNRAS, 295, 691
\reference{} Bonnell, I.A., \& Clarke, C.J. 1999, MNRAS, 309, 461
\reference{} Bresolin, F., \& Kennicutt, R. C., Jr. 1997, AJ, 113, 975
\reference{} Bromm, V., Coppi, P.S., \& Larson, R.B. 1999, ApJ, 527, 5
\reference{} Brown, A. G. A. 1998, in The Stellar Initial Mass Function, 
eds. G. Gilmore \& D. Howell, ASP Conf. Ser., 142, p. 45
\reference{} Calzetti, D. 1997, AJ, 113, 162
\reference{} Cayrel, R. 1990, in Physical Processes in
Fragmentation and Star Formation, ed. R. Capuzzo-Dolcetta, C.
Chiosi \& A. Di Fazio, (Dordrecht: Kluwer),  p. 343
\reference{} Chabrier, G., \& Mera, D. 1997, A\&A, 328, 83 
\reference{} Charlot, S., Ferrari, F., Matthews, G. J., \& Silk, J.
1993, ApJ, 419, L57
\reference{} Clarke, C. 1998, in The Stellar Initial Mass Function, 
eds. G. Gilmore \& D. Howell, ASP Conf. Ser., 142, p. 189
\reference{} Comer\'on, F., Rieke, G. H., Burrows, A., \& Rieke, M. J. 1993, ApJ, 416, 185 
\reference{} Comer\'on, F., \& Torra, J. 1996, A\&A, 314, 776  
\reference{} D'Antona, F., \& Mazzitelli, I. 1994, ApJS, 90, 467
\reference{} De Marchi, G., \& Paresce, F. 1997, ApJ, 476, L19
\reference{} De Marchi, G., Paresce, F., \& Pulone, L. 2000, ApJ, 530, 342
\reference{} Devereux, N. A., 1989, ApJ, 346, 126
\reference{} Doane, J.S., \& Matthews, W.G. 1993, ApJ, 419, 573
\reference{} Doyon, R., Joseph, R.D., \& Wright, G.S. 1994, ApJ, 421, 101
\reference{} Efremov, Y.N., \& Elmegreen, B.G. 1998, MNRAS, 299, 588
\reference{} Elmegreen, B.G. 1983, MNRAS, 203, 1011
\reference{} Elmegreen, B.G. 1993, ApJ, 419, L29
\reference{} Elmegreen, B.G. 1997, ApJ, 486, 944 
\reference{} Elmegreen, B.G. 1998, in Unsolved Problems in
Stellar Evolution, ed.  M.  Livio, 
(Cambridge: Cambridge University Press), p. 59
\reference{} Elmegreen, B.G. 1999a, ApJ, 515, 323
\reference{} Elmegreen, B.G. 1999b, ApJ, 522, 915
\reference{} Elmegreen, B.G. 1999c, in 
The evolution of Galaxies on Cosmological Timescales,
ed. J.E. Beckman \& T.J. Mahoney, ASP Conf. Ser., 187, p.145 
\reference{} Elmegreen, B.G. 2000a, MNRAS, 311, L5 
\reference{} Elmegreen. B.G. 2000b, ApJ, 530, 277 
\reference{} Elmegreen, B.G. 2000c, ApJ, 539, in press, astro-ph/0005455
\reference{} Elmegreen, B.G. 2000d, in Star Formation from the Small to
the Large Scale, eds. F. Favata, A.A. Kaas, \& A. Wilson, ESA
Publications Division at ESTEC, Noordwijk, Netherlands, ESA SP-445, in press,
astro-ph/0005189
\reference{} Elmegreen, B.G., Efremov, Y.N., Pudritz, 
R., \& Zinnecker, H. 2000e, in Protostars and Planets IV, eds. 
V. G. Mannings, A. P. Boss, \& S. S. Russell, (Tucson: Univ. Arizona Press), 
p. 179
\reference{} Elmegreen, B. G., \& Mathieu, R 1983, ApJ, 202, 305
\reference{} Elmegreen, B.G., \& Efremov, Yu.N. 1996, ApJ, 466, 802
\reference{} Elmegreen, B.G., \& Falgarone, E. 1996, ApJ, 471, 816
\reference{} Elmegreen, B.G., \& Efremov, Yu.N. 1997, ApJ, 480, 235
\reference{} Falgarone, E., Puget, J.L., \& P\'erault, M. 1992, A\&A, 257, 715
\reference{} Fabian, A.C. 1994, ARAA, 32, 277
\reference{} Feinstein, C. 1997, ApJS, 112, 29
\reference{} Fischer, P., Pryor, C., Murray, S., Mateo, M., \& Richtler, T. 1998, AJ, 115, 592
\reference{} Fleck, R.C., Jr. 1996, ApJ, 458, 739
\reference{} Freedman, W.L. 1985, ApJ, 299, 74
\reference{} Garmany, C.D., Conti, P.S., \& Chiosi, C. 1982, ApJ, 263, 777
\reference{} Giersz, M., Heggie, D.C. 1996, MNRAS, 279, 1037
\reference{} Greggio, L., Marconi, G., Tosi, M., \& Focardi, P. 1993, AJ, 105, 894 
\reference{} Grillmair, C.J., et al. 1998, AJ, 115, 144
\reference{} Heithausen, A., Bensch, F., Stutzki, J., Falgarone, E. \& Panis, J. F. 1998, A\&A, 331, 65
\reference{} Hill, J.K., Isensee, J.E., Cornett, R.H.,
Bohlin, R.C., O'Connell, R.W., Roberts, M.S., Smith, A.M., \& Stecher, T.P. 1994,
ApJ, 425, 122
\reference{} Hill, R.S., Cheng, K.-P., Bohlin, R.C., O'Connell, R.W.,
Roberts, M.S., Smith, A.M., \& Stecher, T.P. 1995, ApJ, 446, 622
\reference{} Hillenbrand, L.A., \& Hartmann, L. 1998, ApJ, 492, 540
\reference{} Hillenbrand, L.A., \& Carpenter, J.M. 2000, ApJ, in press, astroph/0003293
\reference{} Holtzman, J. A. et al. 1997, AJ, 113, 656
\reference{} Holtzman, J. A., Watson, A. M., Baum, W.A., Grillmair, C.J., Groth, E.J., 
Light, R.M., Lynds, R., O'Neil, E.J., Jr. 1998, AJ, 115, 1946
\reference{} Humphreys, R.M., \& McElroy, D.B. 1984, ApJ, 284, 565
\reference{} Hunter, J. H., Jr., \& Fleck, R. C., Jr. 1982, ApJ, 256, 505
\reference{} Hunter, D.A., Light, R.M., Holtzman, J.A., Lynds, R.,
O'Neil, E.J., Jr., Grillmair, C.J. 1997, ApJ, 478, 124
\reference{} Jones, B. F., \& Walker, M. F. 1988, AJ, 95, 1755
\reference{} Kaas, A.A., \& Bontemps, S.
2000, in From Darkness to Light: Origin and Early
Evolution of Young Stellar Clusters, eds. T. Montmerle \& P. Andr\'e, 
ASP Conf. Series, in press
\reference{} Kennicutt, R.C., Edgar, B.K., \& Hodge, P.W.
1989, ApJ, 337, 761
\reference{} Kennicutt, R.C., Jr., Tamblyn, P., \& Congdon, C.W. 1994, ApJ, 435, 22
\reference{} Kontizas, E., Xiradaki, E., \& Kontizas, M. 1989, Ap.Sp.Sci., 156, 81
\reference{} Kronberg, P. P., Biermann, P., \& Schwab, F. R. 1985, ApJ, 291, 693
\reference{} Kroupa, P., Tout, C.A., \& Gilmore, G. 1990, MNRAS, 244, 76
\reference{} Lada, E. A., \& Lada, C. J. 1995, AJ, 109, 1682
\reference{} Langer, W.D., Velusamy, T., Kuiper, T.B.H., Levin, S., Olsen,
E., \& Migenes, V. 1995, ApJ, 453, 293
\reference{} Larson, R.B. 1973, MNRAS, 161, 133
\reference{} Larson, R.B. 1978, MNRAS, 184, 69
\reference{} Larson, R.B. 1981, MNRAS, 194, 809
\reference{} Larson, R.B. 1982, MNRAS, 200, 159
\reference{} Larson, R.B. 1992, MNRAS, 256, 641
\reference{} Larson, R.B. 1998, MNRAS, 301, 569
\reference{} Leitherer, C. 1999, in Galaxy Interactions at Low and High 
Redshift, 
IAU Symp. 186, eds. J. E. Barnes \& D. B. Sanders, 
(Kluwer: Dordrecht), p.243
\reference{} Lejeune, C., \& Bastien, P. 1986, ApJ, 309, 167
\reference{} Loewenstein, M., \& Mushotzky, R. F. 1996, ApJ, 466, 695
\reference{} Lu, L., Sargent, W., Churchill, C., \& Vogt, S. 1996, ApJS, 107, 475
\reference{} Luhman, K. L., Rieke, G. H., Lada, C. J., \& Lada, E. A. 1998, ApJ, 508, 347
\reference{} Luhman, K. L., \& Rieke, G. H. 1998, ApJ, 497, 354
\reference{} Luhman, K. L., \& Rieke, G. H. 1999, ApJ, 525, 440 
\reference{} Maeder, A. 1985, A\&A, 147, 300
\reference{} Marconi, G., Tosi, M., Greggio, L., \& Focardi, P. 1995, AJ, 109, 173
\reference{} Massey, P. 1998, in The Stellar Initial Mass Function, 
eds. G. Gilmore \& D. Howell, ASP Conf. Ser., 142, p. 17
\reference{} Massey, P., Johnson, K.E., \& DeGioia-Eastwood, K. 1995a, ApJ, 454, 151
\reference{} Massey, P., Lang, C.C., DeGioia-Eastwood,
K., \& Garmany, C.D. 1995b, ApJ, 438, 188
\reference{} Massey, P., \& Hunter, D.A. 1998, ApJ, 493, 180
\reference{} McKee, C.F., \& Williams, J.P. 1997, 476, 144
\reference{} Motte, F., Andr\'e, P., \& Neri, R. 1998, A\&A, 336, 150
\reference{} Muench, A.A., Lada, E.A., \& Lada, C.J. 2000, ApJ, 533, 358
\reference{} Murray, S.D., \& Lin, D.N.C. 1996, ApJ, 467, 728
\reference{} Myers, P. C. 2000a, ApJ, 530, L119
\reference{} Myers, P. C. 2000b, in From Darkness to Light: Origin and Early
Evolution of Young Stellar Clusters, eds. T. Montmerle \& P. Andr\'e, 
ASP Conf. Series, in press
\reference{} Nakano, T., Hasegawa, T., \& Norman, C. 1995, ApJ, 450, 183
\reference{} Najita, J.R., Tiede, G.P., \& Carr, J.S. 2000, ApJ, in press
\reference{} Nissen, P., Gustafsson, B., Edvardsson, B., \& Gilmore, G. 1994, A\&A, 285, 440
\reference{} Norberg, P., \& Maeder, A. 2000, A\&A, 359, 1025
\reference{} Oey, M. S., \& Clarke, C. J. 1998, AJ, 115, 1543
\reference{} Padoan, P., Nordlund, A., \& Jones, B. J. T. 1997, MNRAS, 288, 145
\reference{} Pandey, A. K., Mahra, H. S., \& Sagar, R. 1992, Astr.Soc.India, 20, 287
\reference{} Parker, J.W., Hill, J.K., Cornett, R.H., Hollis, J., Zamkoff, E., Bohlin, R.C.,
O'Connell, R.W., Neff, S.G., Roberts, M.S., Smith, A.M., \& Stecher, T.P.
1998, AJ, 116, 180
\reference{} Paresce, F., \& De Marchi, G. 2000, ApJ, 534, 870
\reference{} Pichardo, B,. V\'azquez-Semadeni, E., Gazol, A., Passot, T., 
\& Ballesteros-Paredes, J. 2000, ApJ, 532, 353
\reference{} Price, N. M., \& Podsiadlowski, Ph. 1995, MNRAS, 273, 1041
\reference{} Pulone, L., de Marchi, G., \& Paresce, F. 1999, A\&A, 342, 440
\reference{} Rees, M.J. 1976, MNRAS, 176, 483
\reference{} Renzini, A., Ciotti, L., D'Ercole, A., \& Pellegrini, S. 1993, ApJ, 419, 52
\reference{} Rieke, G.H., Lebofsky, M.J., Thompson, R.I., Low, F.J.,
\& Tokunaga, A.T. 1980, ApJ, 238, 24
\reference{} Rieke, G.H., Loken, K., Rieke, M.J., Tamblyn, P. 1993, AJ, 412, 99
\reference{} Salpeter, E.E. 1955, ApJ, 121, 161
\reference{} Sagar, R., Piskunov, A. E., Miakutin, V. I., \& Joshi, U. C.
1986, MNRAS, 220, 383
\reference{} Sagar, R., \& Bhatt, H. C. 1989, J. Astrop.Astron., 10, 173
\reference{} Sagar, R., \& Richtler, T. 1991, A\&A, 250, 324
\reference{} S\'anchez D., N.M., \& Parravano, A. 1999, ApJ, 510, 795
\reference{} Satyapal, S., et al. 1995, ApJ, 448, 611
\reference{} Satyapal, S., et al. 1997, ApJ, 483, 148
\reference{} Scalo, J. M. 1985, in Protostars and Planets II,
ed. D. C. Black, \& M. S. Matthews, (Tucson: Univ. of Arizona), p. 201
\reference{} Scalo, J.S. 1986, Fund.Cos.Phys, 11, 1
\reference{} Scalo, J.S. 1998, in The Stellar
Initial Mass Function, eds. G. Gilmore \& D. Howell, ASP Conf. Ser., 142, p. 201 
\reference{} Schaerer, D. 1996, ApJ, 467, L17
\reference{} Schroeder, M. C., \& Comins, N. F. 1988, ApJ, 326, 756
\reference{}Selman, F., Melnick, J., Bosch, G., \& Terlevich, R. 1999, A\&A, 347, 532
\reference{} Shu, F. H., Adams, F. C., \& Lizano, S. 1987, ARAA, 25, 23
\reference{} Silk, J. 1995, ApJ, 438, L41
\reference{} Silk, J. 1977, ApJ, 214, 718
\reference{} Silk, J., \& Takahashi, T. 1979, ApJ, 229, 242
\reference{} Sirianni, M., Nota, A., Leitherer, C., de Marchi, G., \& Clampin, M.
2000, ApJ, 533, 203
\reference{} Smith, D.A., Herter, T., Haynes, M.P., Beichman, C.A., \&
Gautier, T.N., III 1995, ApJ, 439, 623
\reference{} Smith, D. A., Herter, T., \& Haynes, M. P. 1998, ApJ, 494, 150
\reference{} Stasi\'nska,G., \& Leitherer, C. 1996, ApJS, 107, 427
\reference{} Strom, K. M., Kepner, J., \& Strom, S. E. 1995, ApJ, 438, 813 
\reference{} St\"utzki, J., Bensch, F., 
Heithausen, A., Ossenkopf, V., \& Zielinsky, M. 1998, A\&A, 336, 697
\reference{} Subramaniam, A., Sagar, R., \& Bhatt, H. C. 1993, A\&A, 273, 100
\reference{} Tachihara, K., Hara, A., Obayashi, A., Yonekura, Y., 
Onishi, T., Mizuno, A., \& Fukui, Y. 2000, in From Darkness to Light: Origin and Early
Evolution of Young Stellar Clusters, eds. T. Montmerle \& P. Andr\'e, 
ASP Conf. Series, in press
\reference{} Testi, L., \& Sargent, A.I. 1998, ApJ, 508, L91
\reference{} Testi, L., Palla, F., \& Natta, A. 1999, A\&A, 342, 515
\reference{} Tohline, J. E. 1980, ApJ, 239, 417
\reference{} Vazquez, R. A., Baume, G., Feinstein, A., \& Prado, P. 1996, A\&AS, 116, 75
\reference{} Verschuur, G. 1993, AJ, 106, 2580
\reference{} Wang, B., \& Silk, J. 1993, ApJ, 406, 580
\reference{} Wright, G.S., Joseph, R.D., Robertson, N.A., James, P.A.,
\& Meikle, W.P.S. 1988, MNRAS, 233, 1
\reference{} Wyse, R.F.G. 1997, ApJ, 490, L69
\reference{} Wyse, R.F.G. 1998, in The Stellar Initial Mass Function, 
eds. G. Gilmore \& D. Howell, ASP Conf. Ser., 142, p. 89
\reference{} Yoshii, Y., \& Saio, H. 1985, ApJ, 295, 521
\reference{} Zinnecker, H. 1984, MNRAS, 210, 43
\reference{} Zinnecker, H. 1986, in Luminous Stars and Associations in
Galaxies, IAU Symposium 116, eds.  C.W.H. de Loore, A.J. Willis, P. Laskarides,
(Dordrecht: Reidel), p. 271
\reference{} Zinnecker, H. 1989, in Evolutionary Phenomena in Galaxies, ed.                  
J.E. Beckman \& B.E.J. Pagel, (Cambridge: Cambridge Univ. Press), p. 113   
\end{references}
\end{document}